\documentclass[conference]{IEEEtran}
\IEEEoverridecommandlockouts
\usepackage{cite}
\usepackage{amsmath,amssymb,amsfonts}
\usepackage{algorithmic}
\usepackage{graphicx}
\usepackage{textcomp}
\usepackage{xcolor}
\usepackage[colorinlistoftodos,prependcaption]{todonotes}

\usepackage{orcidlink}
\newcommand\orcid[1]{\orcidlink{#1}~\text{#1}}

\makeatletter
\newcommand{\linebreakand}{%
  \end{@IEEEauthorhalign}
  \hfill\mbox{}\par
  \mbox{}\hfill\begin{@IEEEauthorhalign}
}
\makeatother

\def\BibTeX{{\rm B\kern-.05em{\sc i\kern-.025em b}\kern-.08em
    T\kern-.1667em\lower.7ex\hbox{E}\kern-.125emX}}
\begin{document}

\title{F*** workflows: when parts of FAIR are missing \\
\thanks{Notice: This manuscript has been authored by UT-Battelle, LLC under Contract No. DE-AC05-00OR22725 with the U.S. Department of Energy. The publisher, by accepting the article for publication, acknowledges that the U.S. Government retains a non-exclusive, paid up, irrevocable, world-wide license to publish or reproduce the published form of the manuscript, or allow others to do so, for U.S. Government purposes. The DOE will provide public access to these results in accordance with the DOE Public Access Plan (http://energy.gov/downloads/doe-public-access-plan).}
}

\author{
\IEEEauthorblockN{Sean R. Wilkinson}
\IEEEauthorblockA{\textit{Oak Ridge Leadership Computing Facility} \\
\textit{Oak Ridge National Laboratory}\\
Oak Ridge, TN, USA \\
\orcid{0000-0002-1443-7479}}
\and
\IEEEauthorblockN{Greg Eisenhauer}
\IEEEauthorblockA{\textit{College of Computing} \\
\textit{Georgia Institute of Technology}\\
Atlanta, GA, USA \\
\orcid{0000-0002-2070-043X}}
\and
\IEEEauthorblockN{Anuj J. Kapadia}
\IEEEauthorblockA{\textit{Computational Sciences and Engineering} \\
\textit{Oak Ridge National Laboratory}\\
Oak Ridge, TN, USA \\
\orcid{0000-0003-2755-4495}}
\linebreakand
\IEEEauthorblockN{Kathryn Knight}
\IEEEauthorblockA{\textit{Oak Ridge Leadership Computing Facility} \\
\textit{Oak Ridge National Laboratory}\\
Oak Ridge, TN, USA \\
\orcid{0000-0003-2976-0049}}
\and
\IEEEauthorblockN{Jeremy Logan}
\IEEEauthorblockA{\textit{Computer Science and Mathematics} \\
\textit{Oak Ridge National Laboratory}\\
Oak Ridge, TN, USA \\
\orcid{0000-0003-1529-3048}}
\and
\IEEEauthorblockN{Patrick Widener}
\IEEEauthorblockA{\textit{Oak Ridge Leadership Computing Facility} \\
\textit{Oak Ridge National Laboratory}\\
Oak Ridge, TN, USA \\
\orcid{0000-0002-5882-0816}}
\linebreakand
\IEEEauthorblockN{Matthew Wolf}
\IEEEauthorblockA{\textit{Computer Science and Mathematics} \\
\textit{Oak Ridge National Laboratory}\\
Oak Ridge, TN, USA \\
\orcid{0000-0002-8393-4436}}
}

\maketitle


\begin{abstract}

The FAIR principles for scientific data (Findable, Accessible, Interoperable, Reusable) are also relevant to other digital objects such as research software and scientific workflows that operate on scientific data. The FAIR principles can be applied to the data being handled by a scientific workflow as well as the processes, software, and other infrastructure which are necessary to specify and execute a workflow. The FAIR principles were designed as guidelines, rather than rules, that would allow for differences in standards for different communities and for different degrees of compliance. There are many practical considerations which impact the level of FAIR-ness that can actually be achieved, including policies, traditions, and technologies. Because of these considerations, obstacles are often encountered during the workflow lifecycle that trace directly to shortcomings in the implementation of the FAIR principles. Here, we detail some cases, without naming names, in which data and workflows were Findable but otherwise lacking in areas commonly needed and expected by modern FAIR methods, tools, and users. We describe how some of these problems, all of which were overcome successfully, have motivated us to push on systems and approaches for fully FAIR workflows.

\end{abstract}



\section{Introduction}

The FAIR principles -- Findable, Accessible, Interoperable, and Reusable -- represent a set of guidelines for management and stewardship of scientific data \cite{fair-principles}. They were designed and endorsed by a diverse set of stakeholders because of the urgent need to improve the infrastructure supporting the reuse of scholarly data. Additionally, the FAIR Principles put specific emphasis on enhancing the ability of machines to automatically find and use the data, in addition to supporting its reuse by human scholars. This was not common among other similar data initiatives. 

The FAIR principles chiefly concern metadata, which means that they can be applied to data as well as anything that can be described by data. For example, FAIR can be applied to the data being handled by a scientific workflow as well as to software and other infrastructure necessary to specify and execute that workflow. In fact, the application of the FAIR principles beyond scientific data is an active research topic in areas such as research software \cite{research-software}, computational workflows \cite{goble2020}, and even Jupyter notebooks \cite{fair-notebook}. There are needs to continue to adapt research data management to make data more machine-actionable. Methods for increasing automation, such as model-driven approaches, promise better efficiency while reducing sources of human-generated errors, helping to promote reuse and reproducibility of science data \cite{loganbtsd2021}. 
Because the FAIR principles were designed for broad applicability, they are deliberately non-prescriptive, because the meaning of ``necessary and sufficient metadata'' varies from one domain to the next. As a result, they are more like guidelines than actual rules, and different communities have different standards for what it means to be FAIR-compliant.

Many scientists do agree, however, that ``Science is better when it adopts FAIR'' \cite{caw2021-report}, and as a result, FAIR has grown from its initial publication into an international movement, but it has not been without growing pains. FAIR is not prescriptive, which means that there is no checklist to follow when working toward compliance, and some of the principles are easier to implement than others. Naturally, there are also practical considerations like Return on Investment and cost optimization that must be taken into account \cite{turning-fair-into-reality}.

Here, we detail some cases, without naming names, in which data and workflows were Findable but otherwise lacking in the other areas commonly needed and expected by modern FAIR methods, tools, and users. We describe how some of these problems have motivated us to push on systems and approaches for fully FAIR workflows. Fortunately, these examples are compiled from projects that ultimately completed successfully, but where FAIR-related problems led to delays, confusion, or lost opportunities.

As a convention, we will simply use ``data'' in the text rather than ``(meta)data'', ``meta/data'', or other shorthand notations which might distract the reader from our purpose. We focus here on both data consumed and produced by workflows as well as a broad variety of differently structured metadata about those workflows and their data. All of this metadata is also data in its own right, of course, and for simplicity, we will simply use ``data'' to refer to all of it.

We emphasize that our purpose here is to draw attention to the practical realities involved in creating different types of shared scientific data, software, and workflows. Such efforts are difficult enough in a world of competing and sometimes contradictory priorities and constraints. Identifying the people and organizations that have shared their science with us does not serve that purpose, and thus we have obscured those names in the following sections. Our personal experiences are properly used as a lens to see opportunities more clearly for a better ecosystem.


\section{Missing Letters}

The level of adherence to the FAIR principles varies across implementations, and here we relate our experiences when different letters of FAIR are ``missing''. In other words, we describe examples of obstacles we have encountered, grouped in terms of the FAIR principle which was most lacking. All of these examples satisfied expectations for what FAIR considers Findable, but each shows weaknesses in satisfying criteria for being Accessible, Interoperable, or Reusable.

\subsection{Accessible}

The first step in using data is, of course, to find them, and having found them, the user needs to know how they can be accessed, possibly including authentication and authorization. To that end, there are two FAIR sub-principles regarding Accessibility. The first is that the data are retrievable by their identifier using a standardized communications protocol which is open, free, universally implementable, and able to support authentication and authorization procedures, where necessary. The second is that metadata are Accessible even when the data are no longer available. The example we describe here exemplifies a situation in which, according to the FAIR principles, the data are available and seem to fit the definition for being Accessible, but they still are not machine-actionable in the ways that the FAIR principles were intended to enable.

While gathering data for this project, we encountered a variety of Accessibility issues, many of which likely resulted from good intentions that had unforeseen consequences. One frequent hindrance involved cases where the data, though available, was not readily processable because the authors had favored making data human-readable at the expense of making it machine-actionable. As a result, we encountered tables on web pages that were not easily disentangled from the markup used to structure and style the page. Another case included data that could only be downloaded as a PDF, and the data were presented there in tabular form. Both of these cases are representative of the same difficult problem, namely that there is no simple way to infer structure from these visual presentation modes. Even if data elements can be extracted, the tabular structure is lost, and so leveraging such a dataset comes with the cost of manual extraction to a structured format such as a spreadsheet or a CSV file.

One dataset, which we intentionally refrain from identifying here, offered a sort of ``perfect storm'' of Accessibility challenges. First, as the data required a Data Usage Agreement to be in place, access to the data was tightly restricted. We were given a password to access an FTP server and instructed to use a particular client (Filezilla). Indeed, we found that other FTP clients were blocked, so access required the additional step of downloading and installing Filezilla rather than simply using an already-installed client. 

Once we were able to connect to the FTP site, we found we could see not only our data, but also datasets that had been prepared for others. Keeping our eyes on our own work, we downloaded our designated directory, which contained roughly 50 sub-directories, each with a separate encrypted, password-protected file. The encryption mechanism turned out to be specific to Microsoft Windows; we therefore had to move the data to a separate Windows machine, decrypt each of the files separately, and then move the entire file tree back to the Linux machine where the data would be processed.

Processing the data continued to present challenges. First, the naming scheme across the sub-directories' files, which had been collected over a number of years, was not consistent. This made scripting more difficult, as we needed to provide an explicit mapping from the underlying structure to the actual filenames.

As a final challenge, the individual data files had been encoded using a fixed-width format which required a custom reader to be created just to load and understand the content of the files. Fortunately, the required information to implement the reader was provided, but it required reading through a significant portion of the somewhat lengthy documentation. This stood out in contrast to other datasets that were being gathered for the project, which were mostly CSV, netCDF, and Microsoft Excel files; none of the others provided such a steep barrier to entry.

\subsection{Interoperable}

Data usually need to be integrated with other data. In addition, data need to interoperate with applications or workflows for analysis, storage, and processing. For this reason, there are three FAIR sub-principles that target Interoperability. First, the data should use a formal, Accessible, shared, and broadly applicable language for knowledge representation. Second, the data should use vocabularies that also follow FAIR principles. Third, the data should include qualified references to other data.

One solution posed to enable Interoperability of data is to remodel data according to a concept model or ontology, and then either to map that data into an existing system or to build a new system that can leverage this new model, e.g., \cite{queralt2022}. However, there are some difficulties with this solution, as it partially assumes that the data at hand easily map to a data model, both conceptually and in terms of the labels provided for the data. It also assumes that the original data are labeled in such a way that a person or group of persons can translate them into broadly applicable language that uses FAIR vocabularies and includes qualified references.

In our experiences, invisible context abounds in data stored in various institutional data silos, especially regarding policy changes which may have seemed ``obvious'' at the time and therefore not worth recording. One concrete example is the complete set of telemetry data from a decommissioned computer system. During the system's active lifetime, all contextual information about what the telemetry messages meant was contained in the system's manual, or ``man'' pages. Once the system disappeared, so did the man pages, along with all of the relevant context about how the messages in the various telemetry files connected with one another and, crucially, if they would pose any security risks should they be made available outside of the institution. Today, the data, which are comprised of approximately 3 terabytes of mostly unstructured text files, exist in a kind of limbo, where they cannot be destroyed but also cannot be shared widely, since the cyber security staff cannot say with certainty that the data are safe to release to the wider public. Because crucial contextual information about the requisite qualified metadata references between the files has been lost with the decommissioning of the system, there is no way to render this dataset Interoperable with itself or other systems. 

Another tricky facet of Interoperability is the means of knowledge representation, which is asserted in the first Interoperability principle. Both ``shared'' and ``broadly applicable'' seem like reasonable conditions for language at first. Instead, it turns out that this is actually quite complicated because domains often disagree within themselves about what is ``shared'' and/or ``broadly applicable.'' Star and Griesemer \cite{star1989} describe the heterogeneity inherent in scientific work, even within one domain, and develop the concept of the \textit{boundary object}, where a single object can be used for different purposes by different people, even by people working toward a similar research goal.

Crucially, the concept of a boundary object is predicated on standardization, where something with a standardized or predetermined structure, such as a map or a bird, can maintain a common identity, but may have different conceptual meanings in different communities of practice. In a later paper clarifying how the concept of a boundary object has evolved in the research community, Star underscores that scientists are able to cooperate even when consensus is rarely reached on how to describe a boundary object \cite{star2010}. 

For instance, one concrete example appears in medical procedures within the domain of a hospital, where there may be both local and shared meaning about these procedures. Local meaning might be the work of nursing staff, where perceived granularity in various procedures (captured in the Nursing Interventions Classification, or NIC) differs with what might be coded as a Current Procedural Terminology (CPT) code. For instance, the NIC includes the category of ``Spiritual Support'' (5420). This same concept does show up in CPT codes, but is specific to the work of Chaplains (Q9001-Q9003). If a hospital system opted for CPT codes as the ``broadly applicable'' \emph{lingua franca} for procedures, this granular concept for the specific nature of nursing work is completely lost, as there is only one CPT code for nursing care in a hospital (99211). The boundary object, ``nursing care'', is understood both by nurses and the rest of the hospital staff, but it is conceptually different, and therefore rendered differently, depending on what classification system codes this object. Using terms like ``shared'' and/or ``broadly applicable'' ignores that subgroups within a domain may need to maintain a vague identity for a common object, and for this reason obscures how this principle can erase potentially essential granularity in how concepts get represented. This, in turn, affects Interoperability with systems that use classifications of varying granularity.

A recent study describes the roadblocks encountered when a clinical research team attempted to retrieve 23,186 abdominal CT exams from radiology systems \cite{magudia2021}. One of the first difficulties they encountered were the Accession Numbers (ACCs) used across different hospitals, which were not only inconsistent from one system to the next, but also linked to resulting images differently ``such that the images were most often linked to the abdomen ACC, but they could also be linked to the pelvis ACC, or even the chest ACC if chest, abdomen, and pelvis CT exams were acquired together. These linkages varied over time due to changing systems and policies.''

It is that final sentence which reveals the tremendous difficulty in applying the third Interoperability principle to this example. It is unclear exactly how changing systems and policies should be handled or described via qualified references to other data. In fact, their discussion of these roadblocks includes evolving policies as a major obstacle to research, because the policies reflected in the data were only understood when the team members were able to find individuals whose memories spanned these policy changes.

\subsection{Reusable}

Ultimately, the goal of FAIR is to optimize the reuse of data. To achieve this goal, metadata and data should be well-described so that they can be replicated and/or combined in different settings. As a result, FAIR also includes a Reusability principle, which somewhat confusingly has one sub-principle comprised of three sub-sub-principles. Broadly speaking, the Reusability principle says that data should be richly described with a plurality of accurate and relevant attributes. More specifically, it says that data should be released with a clear and Accessible data usage license, that data should be associated with detailed provenance, and that data should meet domain-relevant community standards.

There are a number of open problems in workflow science that make it difficult to apply the Reusability principle to scientific workflows themselves. One of these problems is that the word ``workflow'' does not have a consensus community definition \cite{community-summit}. Without community agreement on precisely what constitutes a workflow, there can be no agreement on precisely what metadata are necessary to describe a workflow fully. Goble et al.\ \cite{goble2020} propose two areas in which the FAIR principles apply to workflows: FAIR data both for and from workflows, and criteria for FAIR digital objects. In the case of FAIR data for and from workflows, workflows would include descriptive metadata about the data produced as well as metadata that helps trace that data's provenance. When considering workflows as FAIR digital objects, as in the second case, a workflow is seen as an ``object'' describing methodology that may be subsequently distributed, used, cited, and modified. Without a community agreement on what a workflow ``is'', however, these two areas for applying FAIR remain ill-defined.

For our purposes here, then, in applying the Reusability principle to workflows, we will consider scientific workflows to include all of the data, processes, and infrastructure necessary to follow the steps of the Scientific Method for a given experiment. Reusability obstacles then arise for scientific workflows at several levels.

One source of Reusability applies to tracking the provenance of the data being manipulated by workflows. ``Provenance'' is both subjective and tricky, even within the context of a domain (more on the complications of the word ``domain-relevant'' later). Provenance is often an afterthought or is handled inconsistently -- not a standard practice or priority. Furthermore, standards for recording provenance are not widely known (e.g., the W3C's PROV-O ontology~\cite{w3c-prov-o} or PROV-JSON serialization~\cite{w3c-prov-json}). For instance, some researchers lack support for post-experiment metadata documentation, resulting in ad-hoc result management like naming output files with experiment parameter values; this is, in fact, a practice we have encountered among colleagues and collaborators working in certain domains at Oak Ridge National Laboratory (ORNL). There is great need for institutional support for systems that provide a convenient means to record domain-specific experimental metadata like inputs, outputs, and context of processes linked to a given workflow. Without such support, the burden then falls on the researcher to invest time into recording this information somewhere and either assuming the technical debt of maintaining a system-of-record for this provenance, or, more practically, using whatever is reasonably available at the time, such as a file or directory name.

Provenance is further complicated when data are aggregated from external sources, where information related to collection methods, coding standards and vocabularies (whether local or more widely-used), modification history, etc.\ is entirely up to an external group or organization to document and make available. Several sizeable datasets stewarded by ORNL arrive from external organizations containing their own local representations that, when aggregated, show inconsistencies at the data level that cannot be clearly traced to a singular process or set of documentation practices (e.g., variable coding standards for medical laboratory tests).

As mentioned earlier, ``domain-relevant'' is also a complicated assertion which is also related to the discussion of boundary objects in the Interoperability section. What, specifically, is a domain? What are the boundaries of a domain, and how are they determined? Beyond finding a clear way to define a domain, many concepts within domains are hotly debated by researchers, so pointing to a standard is essentially meaningless in some cases. Alternatively, they would at least require the ability to record two simultaneous standards for certain concepts depending on the scientific stance of the researcher. An example here would be that they share the same ontology, such as the ontic versus epistemic views of quantum physics, or perhaps asking researchers in electron microscopy to clarify the difference between ``high-tension'' and ``acceleration voltage'', which may or may not be used interchangeably, depending on whom you ask.

Another set of obstacles we have encountered while applying Reusability to workflows appears in the important step of packaging the workflow for consumption by others, such as for reproducing the experiment. Specifically, it is difficult if not impossible to package a workflow for reuse when it contains unwritten, human-centric operations. As a concrete example, we have encountered a workflow in which one particular step involves column-wise pasting of a large number of individual tabular files into a single large file. This is a simple concatenation that could be accomplished with the UNIX ``paste'' command in theory, but the datasets are generally too large for a single paste to suffice without significant performance degradation. The solution requires a series of smaller pastes to be performed over subsets of files until a final paste can merge the pasted subsets. This requires careful planning from a human in order to divide the pasting into parallelizable sub-jobs that will each have a reasonably short runtime and avoid filesystem bottlenecks from working with too large a number of files simultaneously. The scientist also must monitor the jobs after launching them to make sure that they are completing successfully and to keep track of remaining jobs. These tasks can be accomplished by modern scripting languages, but scientists often forego automation for various reasons. How, then, can this workflow be packaged, if so much of it was unwritten and performed by a human?

\section{Discussion}

These examples are compiled from issues and projects that proceeded to success, but where the lack of, or idiosyncratic interpretation of, the FAIR principles has led to delays, confusion, or lost opportunities. These sorts of scenarios are almost assuredly not a surprise to most practitioners in the sciences; in fact, it is generally considered a solved problem. That solution, unfortunately, is to throw people at it -- to hire a graduate student, a post-doc, or a technical data engineer, for example, and just tell them to go away and work on it for six months. Our purpose in revisiting and talking about these problems is as a reminder that it doesn't have to be this way.

Science is ultimately driven not just by the discovery of new information, but also by the communication of that information in a way that it can be reused and built upon and accumulated. With some more attention to the systems and management, we could be generating datasets, software, and workflows that are more Reusable and Accessible in both the context of science and in the context of modern data and cloud architectures. We need to include the human costs and the fragility of the current process in any Return on Investment calculation for developing and adopting new approaches, and in that light, we believe there is a clear case and need for such investment.

The hidden costs of human investment in reuse is known as ``technical debt'' in the software engineering field. Some of us have been working to make this connection between a more generalized concept of technical debt for scientific data and workflows and the need for better and more FAIR infrastructures \cite{reusability-first}. In the broader context, this problem is hardly new or unexpected. It is particularly important now, however, as we see the rise of data-intensive science applications that leverage the explosion in interest in Artificial Intelligence (AI) and Machine Learning (ML) techniques. Feeding such algorithms so that they can generate reasonable, unbiased, scientifically valid results means that the Reusability, Interoperability, Accessibility, and Findability of scientific results and the provenance of the processes and workflows used to generate them must all be readily available.

Awareness of the FAIR principles among scientists grows every day, but for adherence and adoption to keep up, incentive systems will need to improve. Research Funding Organizations (RFOs) can be strong drivers, for example, by acting as both the proverbial carrot and the corresponding stick. RFOs can combine their funding requirements and funding policies (``stick'') with guidance and financial support (``carrot'') for researchers to incorporate tools and standards that make their data FAIR while fitting their research objectives \cite{fair-funding}. In this way, RFOs also encourage research institutions to improve support and facilities for their researchers to create FAIR data, because they need to comply with the funder's requirements. As another example, publication of quality-assured datasets with standardized metadata should be rewarded in the same way that research papers in high-impact journals are \cite{health-data-sharing}. The hope in this approach is two-fold: that researchers would be as eager to share their data as they are to publish their results, and that they would prioritize the quality of the published data. This same idea also extends to the publication of software and workflow tools, which are often available only upon request.

Finally, we should emphasize that the research efforts we have discussed have all led to quality scientific contributions, in spite of the issues and problems we have identified. We are excited by the possibilities of a research enterprise in which not only these kinds of efforts are made easier, but also the process of extending them is made more straightforward.

\section{Conclusion}

At the same time that the FAIR principles ``hit a chord'' \cite{first-gen-fair}, workflows have quickly grown and become the default abstraction in large-scale computational science. As a result, the FAIR principles are also becoming more popular as a common vocabulary for discussing and comparing workflows. We observe that the community's understanding of the interplay between FAIR's components -- and how that interplay is best realized in operating system, runtime, and application software -- has lagged FAIR's increased usage as a \emph{lingua franca}. Until that understanding is better developed, situations such as the ones we have described here are likely to continue to arise. In the near term, we hope that discussions like this help scientists avoid similar time-consuming roadblocks in the pursuit of their scientific objectives. We hope that these personal examples of datasets and opportunities where open science could have used a little more FAIR-ness are not read as a condemnation of any one collaborator, provider, or discipline. Instead, for the longer term, we see this as a call to arms for all of us across the community to remember that these sunk costs exist. There is no reason for them to remain an assumed part of business as usual.

\section*{Acknowledgments}

This research used resources of the Oak Ridge Leadership Computing Facility at the Oak Ridge National Laboratory, which is supported by the Office of Science of the U.S. Department of Energy under Contract No.\ DE-AC05-00OR22725.


\bibliographystyle{IEEEtran}
\bibliography{references}

\end{document}